\documentclass[authoryear,5p,times]{elsarticle}

\RequirePackage[T1]{fontenc}
\usepackage[utf8]{inputenc}
\usepackage{tikz}
\usepackage{amsmath}
\usepackage{amssymb}
\usepackage{url}
\usepackage{graphicx}
\RequirePackage{graphicx}
\RequirePackage{mathptmx}      
\RequirePackage{flushend}
\usepackage[]{algorithm2e}
\usepackage{mathtools}

\DeclareMathOperator*{\argmax}{arg\,max}

\makeatletter
\newcommand\footnoteref[1]{\protected@xdef\@thefnmark{\ref{#1}}\@footnotemark}
\makeatother
\journal{Journal Computers \& Security}
\begin{document}
	\begin{frontmatter}
		\title{Enhancing the Insertion of NOP Instructions to Obfuscate Malware via Deep Reinforcement Learning}
		\author[label1]{Daniel Gibert\corref{cor1}}
		\address[label1]{Unversity College Dublin, CEADAR, Belfield Office Park, Dublin, Ireland}
		\cortext[cor1]{I am corresponding author}
		
		\ead{daniel.gibert@ucd.ie}
		\ead[url]{https://scholar.google.com/citations?user=lAAwRpMAAAAJ\&hl}
		
		\author[label2]{Matt Fredrikson}
		\address[label2]{Carnegie Mellon University, Forbes Avenue Pittsburgh, PA 15213, United States of America}
		
		\author[label3]{Carles Mateu}
		\author[label3]{Jordi Planes}
		\address[label3]{University of Lleida, Jaume II, 69, Lleida, Spain}
		\author[label1]{Quan Le}

	\begin{abstract}
		Current state-of-the-art research for tackling the problem of malware detection and classification is centered on the design, implementation and deployment of systems powered by machine learning
		because of its ability to generalize to never-before-seen malware families and polymorphic mutations. However, it has been shown that machine learning models,  in particular deep neural networks, lack robustness against crafted inputs (adversarial examples). 
		In this work, we have investigated the vulnerability of a state-of-the-art shallow convolutional neural network malware classifier against the dead code insertion technique. 
		We propose a general framework powered by a Double Q-network to induce misclassification over malware families. The framework trains an agent through a convolutional neural network to select the optimal positions in a code sequence to insert dead code instructions so that  the machine learning classifier mislabels the resulting executable.
		The experiments show that the proposed method significantly drops the classification accuracy of the classifier to 56.53\% while having an evasion rate of 100\% for the samples belonging to the Kelihos\_ver3, Simda, and Kelihos\_ver1 families. In addition, the average number of instructions needed to mislabel malware in comparison to a random agent decreased by 33\%.
		\begin{keyword}
			Malware Classification \sep  Assembly Language Source Code\sep Obfuscation \sep Reinforcement Learning \sep Deep Q-Network
		\end{keyword}
		
	\end{abstract}	
		
	\end{frontmatter}








\section{Introduction}
Malware is on the rise. Global detections of newly-developed malware keep increasing year after year.~\footnote{\url{https://www.statista.com/statistics/680953/global-malware-volume/}} 
According to AV-TEST Institute,~\footnote{\url{https://www.av-test.org/en/?r=1}} there has been a dramatic spike of new malicious programs and potentially unwanted applications (PUA) year after year, doubling the number of total malware detected from 2015 to 2020. This recent surge of malicious programs is connected to the increasing dependency of people, things and organizations on the Internet, which provides cybercriminals with a vast range of targets to exploit, from traditional personal computers and laptops to industrial systems, mobile phones and Internet of Things (IoT) devices. In addition, the adoption of remote working due to global events such as the current global pandemic has propelled a new rise in cyberattacks, particularly the increase of ransomware attacks~\footnote{\url{https://pages.checkpoint.com/cyber-attack-2021-trends.html}}.


To keep up with malware evolution and be able to mitigate the damage and impact of cyberattacks, it is necessary to constantly improve the computer systems defences.
One essential security element is endpoint protection, i.e. the use of security solutions to protect endpoints or end-user devices from being exploited against zero-day exploits, attacks, data leakages. Within the wide range of tools to secure an endpoint, anti-malware engines are the last layer of defence. More specifically, anti-malware engines are responsible for preventing, detecting and removing malicious software. Traditionally, anti-malware solutions relied on signature-based and heuristic-based methods. However, due to the huge volumes of new malware variants being deployed every day, traditional anti-malware solutions that rely solely on signatures and heuristics manually defined by domain experts cannot keep pace with the rapidly evolving malware. 


During the last decade, research on machine learning (ML) solutions to tackle the problem of malware detection and classification increased because of the ability of machine learning systems to generalize to unseen malware and polymorphic mutations~\citep{Souri2018,GIBERT2020102526,10.1145/3417978}. Although machine learning models~\citep{DBLP:journals/corr/AhmadiGUST15,7847046} and in particular deep neural networks~\citep{DBLP:conf/ccia/GibertBMPSV17,10.1145/3029806.3029823,DBLP:conf/aaai/RaffBSBCN18,krcal2018deep,DBLP:journals/virology/GibertMPV19,8681127} have achieved significant success in the cybersecurity domain, they have been shown to be vulnerable to adversarial examples~\citep{DBLP:journals/corr/GrossePM0M16,DBLP:journals/corr/abs-1901-03583,DBLP:journals/corr/abs-1801-08917} , i.e. modified examples with imperceptible perturbations that cause misclassification at test time.

The attacks presented in the literature mainly focus on minor modifications of the PE header, appending some bytes at the end of sections, inside code caves, at the end of the PE file, modifying the PE Headers, etcetera.
However, these modifications are not usually generated by common obfuscation techniques employed by malware authors in a real world scenario but specifically designed to evade specific ML-based anti-malware engines. On the contrary, malware authors employ a variety of techniques to obfuscate the code and make it more difficult for humans to understand. Common obfuscation techniques are the dead code insertion technique, instruction replacement, code transposition, packing, encryption, among others.

The purpose of this work is to fill this gap. Accordingly, we have analyzed the robustness of a state-of-the-art malware classifier trained on the assembly language instructions of Portable Executables~\citep{DBLP:conf/ccia/GibertBMPSV17,10.1145/3029806.3029823} against the simplest of the obfuscation techniques, the dead code insertion technique. Furthermore, the aforementioned technique has been enhanced with deep reinforcement learning to increase its evasion success rates. To this end, we present a framework that uses Double Q-learning to induce misclassification over malware families. Within this framework, an AI agent is set up to confront the malware classifier and select a sequence of functionality-preserving actions (dead code insertions) to modify the samples. For any given malware sample, the framework can eventually determine the optimal sequence of actions to make the ML-based classifier output an incorrect label. In this study, we have used the Portable Executable files provided by Microsoft for the Big Data Innovators Gathering challenge of 2015 for reproducibility purposes~\citep{DBLP:journals/corr/abs-1802-10135}.

The rest of the paper is organized as follows: Section~\ref{sec:background} introduces the state-of-the-art approaches for malware detection and classification, and the adversarial attacks devised to bypass detection. Section~\ref{sec:rl_framework} describes the proposed framework in detail. Section~\ref{sec:evaluation} presents the experimental setup and the results obtained. Section~\ref{sec:conclusions} summarizes the concluding remarks extracted from this work and presents some future lines of research.

\section{Related Work}
\label{sec:background}
\subsection{Machine Learning for Malware Detection and Classification}
\label{sec:ml_detectors_overview}
Recently, machine learning (ML) has become an appealing signature-less approach for malware detection and classification because of its ability to handle huge volumes of data and to generalize to never-before-seen malware~\citep{GIBERT2020102526,Souri2018,10.1145/3417978}. Research has shifted from traditional approaches based on feature engineering~\citep{DBLP:journals/corr/AhmadiGUST15,7847046} to deep learning approaches~\citep{krcal2018deep,DBLP:conf/ccia/GibertBMPSV17,DBLP:conf/aaai/RaffBSBCN18,AAAI1816133,DBLP:journals/virology/GibertMPV19,8681127,VENKATRAMAN2019377,GIBERT2020101873} because it allows to obviate and replace the time-consuming feature extraction process by an end-to-end system, which typically consists of a neural network with multiple layers, that performs both feature learning and classification altogether. With deep learning, one can start with raw data as features will be automatically learned by the network through training on the labelled data. For instance, 
~\cite{DBLP:conf/ccia/GibertBMPSV17} and~\cite{10.1145/3029806.3029823} presented a shallow convolutional neural network architecture to classify malware based on the assembly language instructions extracted from the assembly language source code of malware from PE executables and Android APKs, respectively. 
Similarly,~\cite{DBLP:conf/aaai/RaffBSBCN18} 
and
~\cite{krcal2018deep} designed a shallow and deep convolutional neural networks, respectively, to detect malware based on the bytes content.
Instead of taking as input the byte sequence, which could consist of a few million time steps,~\cite{AAAI1816133} presented a method for classifying malware by compressing its binary content as a stream of entropy values (or structural entropy) using convolutional neural networks. In other works~\citep{DBLP:journals/virology/GibertMPV19,8681127}, the executables are represented as a grayscale image by interpreting every byte as one pixel in an image, with values ranging from 0 to 255 (0:black, 255:white). Recently, researchers~\citep{VENKATRAMAN2019377,GIBERT2020101873} have started complementing traditional features with features extracted through deep learning. The reason behind this multimodal approach is that each executable includes multiple modalities of information. By only taking as input the raw bytes or opcodes, a lot of information for classification is overlooked. As a result, multiple types of features provide a better abstract representation of the executable characteristics, leading to more accurate ML models.

\subsection{Adversarial Attacks on ML-based Malware Detectors}
Although machine learning has demonstrated impressive performance in several application domains, ranging from computer vision and natural language processing to cybersecurity, ML models have been shown to be vulnerable to adversarial examples~\citep{REN2020346,7467366}, i.e. inputs to the models that an attacker has carefully designed to cause the model to make a mistake, and the domain of cyber security is no exception~\citep{DBLP:journals/corr/GrossePM0M16,DBLP:journals/corr/abs-1901-03583,8553214,DBLP:journals/corr/abs-1801-08917}. 
In addition, in the cyber security domain there really exist actual adversaries, the malware developers, who are strongly motivated to craft their malicious softwares to bypass detection systems in order to steal user information, spread across the network, or perform other hostile activities. Next, some relevant adversarial attack approaches in the literature are presented.


~\cite{DBLP:journals/corr/GrossePM0M16} proposed a gradient-based approach to generate adversarial Android malware examples to evade their own malware detector based on features extracted from (1) permissions and hardware components access requested, (2) API calls made by the application, (3) intents used to communicate with other applications and (4) application components, service, content provider and broadcast receivers used by the applications. They proposed a crafting process that iteratively modifies the feature whose gradient is the largest until the adversarial sample bypasses the detection model.

\cite{8844597} explored the vulnerabilities of byte-based malware detectors  to adversarial malware binaries and presented various strategies to bypass binary detectors by appending a few bytes at the end of the PE headers and sections. Similarly,~\cite{8553214} presented an approach to evade malware detectors by appending a set of carefully-handpicked bytes at the end of the file. In addition,~\cite{DBLP:journals/corr/abs-1901-03583} refined the aforementioned technique by using feature attribution to identify the most influential input features contributing to each decision and thus, reducing the number of bytes needed to be manipulated in order to evade the malware detector.

Alternatively,~\cite{DBLP:journals/corr/abs-1801-08917} presented a pioneering attack using reinforcement learning to craft adversarial examples using a series of actions to manipulate a malware binary. In their work, they trained an agent to evade a malware detector based on features extracted from the PE header and sections metadata, the import and export tables, counts of human readable strings, the byte histogram and a 2D byte-entropy histogram of the executable. To do so, they allowed the agent to perform various mutations on the executable that do not break its file format or alter the code execution, such as appending bytes at the end of sections, adding functions to the import address table. However, as their results show, the evasion rate between their agent and a random agent is practically the same, mainly because  the nature of the mutations implemented, being the use of packing the most successful mutation as it is the technique that affects the most features used for training the ML detector.

The aforementioned attacks~\citep{8844597,8553214,DBLP:journals/corr/abs-1901-03583,DBLP:journals/corr/abs-1801-08917} mainly rely on appending some bytes at the end of the sections or at the end of the PE file. However, these attacks are quite different from those employed by malware authors in a real world scenario to obfuscate the executables. Common obfuscation techniques are the dead code insertion technique, the instruction replacement technique and the subroutine reordering technique, just to name a few. Subsequently, in this work, we address this problem by evaluating the robustness of a state-of-the-art malware classifier~\citep{DBLP:conf/ccia/GibertBMPSV17,GIBERT2021102159} against the simplest obfuscation technique, the dead code insertion technique, and we present a framework to enhance its performance. This classifier has been selected because of its superior performance with respect to deep learning byte-based approaches in the literature~\citep{krcal2018deep,DBLP:conf/aaai/RaffBSBCN18}.

\section{Reinforcement Learning Framework}
\label{sec:rl_framework}
In this paper, we present a Proof of Concept (PoC) of a deep reinforcement learning framework to induce misclassification of malicious Portable Executable (PE) files over malware families.

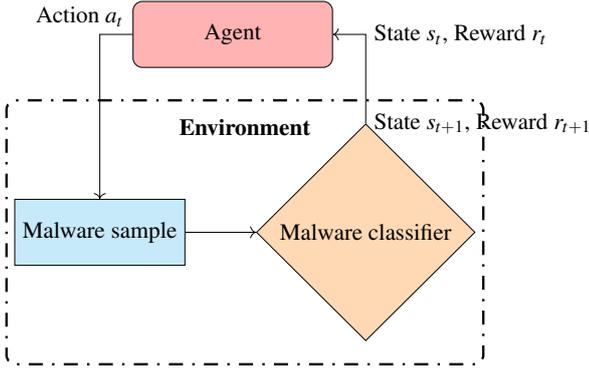
\begin{figure}
	\centering
	\usetikzlibrary{shapes.geometric, arrows.meta, backgrounds, fit, positioning}
	\usetikzlibrary{matrix,chains,scopes,positioning,arrows,fit}
	\tikzstyle{agent} = [rectangle, rounded corners, minimum width=3cm, minimum height=1cm,text centered, draw=black, fill=red!30]
	\tikzset{black dotted mnemonics/.style={draw=black, line width=1pt, dash pattern=on 1pt off 4pt on 6pt off 4pt, minimum width=6cm, minimum height=4cm, rectangle, rounded corners}}
	\tikzstyle{mlw_sample} = [rectangle, minimum width=2cm, minimum height=1cm,text centered, draw=black, fill=cyan!20]
	\tikzstyle{classifier} = [diamond, minimum width=3cm, minimum height=1cm, text centered, draw=black, fill=orange!30]
	\tikzstyle{arrow} = [thick,->,>=stealth]
	\tikzstyle{arrow1} = [thick,<-=stealth]
	\resizebox{0.9\columnwidth}{!}{
		\begin{tikzpicture}
		\node (agent) [agent, xshift=2cm] {Agent};
		\node (sample) [mlw_sample, yshift=-3cm] {Malware sample};
		\node (mlw_classifier) [classifier, right of=sample, xshift=3cm] {Malware classifier};
		\node (environment_component) [black dotted mnemonics, fit= (sample) (mlw_classifier) ] {};
		\node at (environment_component.north) [below, inner sep=3mm] {\textbf{Environment}};
		\draw[->] (sample) -- (mlw_classifier);
		\draw[->] (mlw_classifier.north) node [right] {State $s_{t+1}$, Reward $r_{t+1}$} |- node [right] {State $s_{t}$, Reward $r_{t}$} (agent.east);
		\draw[->] (agent.west) -| node [above, xshift=-0.3cm] {Action $a_{t}$} (sample);
		\end{tikzpicture}
	}
	\caption{Reinforcement learning schema for evading a malware detector. At time step $t$, the agent takes as input the state $s_{t}$ and the reward $r_{t}$. It selects the best action $a_{t}$ to perform, i.e. the location to insert the NOP instruction that has a higher Q-value, and modifies the malware sample. Afterwards, it checks whether or not the mutated sample produces a misclassification. If not, the procedure is repeated until the sample is misclassified or it reaches the maximum number of modifications allowed to be performed by the agent before declaring failure.}
	\label{fig:rl_mlw_environment}
\end{figure}

Reinforcement learning~\citep{10.5555/1622737.1622748} is a branch of machine learning where an agent seeks to learn optimal decision-making by trying to maximize cumulative rewards to achieve a specific goal. The two main components are the environment and the agent. On the one hand, the environment represents the problem to be solved. On the other hand, the agent represents the learning algorithm used to perform actions in the environment in order to maximize the rewards. Thus, reinforcement learning (RL) algorithms study the interaction of agents in such environments and learn to optimize the behavior of the agents in order to maximize the rewards. 
The learning process consists of an agent and an environment that continuously interact with each other, as shown in Figure~\ref{fig:rl_mlw_environment}. At each time step, the agent takes  action $a_{t}$ on the current observation of the environment $s_{t}$, based on its policy $\pi(a_{t}, s_{t})$, and receives a reward $r_{t+1}$ and the next observation of the environment $s_{t+1}$. 

The problem of inducing misclassification of malware samples can be framed as a Makov Decision Proces~\citep{10.2307/24900506}, or MDP, which is a formalism that allows to define the interaction between the agent and environment as a tuple of four elements (S, A, T, R):

\begin{itemize}
	\item $S$: Set of states. At each time step, the state of the environment is an element $s\in S$, where $s_{t}$ denotes the state of the environment at time step $t$. 
	\item $A$: Set of actions. At each time step, the agent chooses an action $a \in A$, where $a_{t}$ denotes the action chosen at time step $t$. The set of actions that can be performed on a particular state $s \in S$, is denoted $A(s)$, where $A(s) \subseteq  A$. Note that in some systems not all actions can be applied in every state. This occurs in our case, as there are positions in the assembly language source code that we will not be able to insert dead instructions because it will break the executable.
	\item $T(s_{t}, a_{t}, s_{t+1})$: State transition function that describes how the environment's state changes when the agent performs an action in a given state. By applying action $a_{t} \in A$ in a state $s_{t} \in S$, the system makes a transition from state $s_{t}$ to a new state $s_{t+1}\in S$, based on a probability distribution over the set of possible transitions. The transition function T is defined as the probability of ending up in state $s_{t+1}$ after performing action $a_{t}$ in state $s_{t}$.
	
	Assuming that the system is Markovian, i.e. the result of an action does not depend on previous actions and previous states, but only depends on the current state, the transition function can be mathematically described as follows:
	\begin{equation}
	T(s_{t}, a_{t}, s_{t+1}) = P(s_{t+1}|s_{t}, a_{t}) = P(s_{t+1}| s_{t}, a_{t}, s_{t-1}, a_{t-1},...)
	\end{equation}

	\item $R(s_{t}, a_{t}, r_{t+1})$: Reward model used to describe the reward value the agent receives from the environment after transitioning from state $s\in S$ to the new state $s^{'} \in S$ with action $a \in A$.
	
\end{itemize}

To sum up, at each time step, the agent will receive some state of the environment $s\in S$. Given this representation, the agent selects an action to take. Then, the environment transitions to a new state and the agent is given a reward as a consequence of its previous actions. This continuous interaction of the agent with the environment creates a trajectory of states, actions and rewards, which can be interpreted as: 
$$ S_{0}, A_{0}, R_{1}, S_{1}, A_{1}, R_{2}, S_{2}, A_{2}, R_{3}, ...$$
Thus, the goal of the agent is to derive a policy $\pi$, the strategy that the agent will pursue, so as to maximize the total amount of rewards it receives over the course of action. In consequence, the agent does not want to maximize the immediate rewards but the cumulative rewards that it will receive over time. At a given time step $t$, the cumulative reward is defined as:
\begin{equation}
G_{t} = \sum_{i=0}^{T-t-1} \gamma^{i} R_{t+i+1}
\end{equation} 
where $T$ is the final time step and $\gamma$ is the discount factor used to control the importance of future rewards. 

The policy $\pi$ is a function that maps a given state $s$ to probabilities of selecting each possible action from that state. In other words, for each state $s\in S$, $\pi$ is a probability distribution over $a \in A(s)$. Most MDPs derive optimal policies by learning value functions. There are two types of value functions: (1) functions of states, or (2) state-action pairs, that estimate how good it is for the agent to be in a given state, or how good it is for the agent to perform a given action in a given state, respectively, where the quality of a state or a state-action pair is given  in terms of expected return. 
Considering that the rewards an agent expects to receive are dependent on the actions it takes in given states and that these are influenced by the policy the agent is following, we can see that the value functions are defined with respect to policies.

So, the state-value function for a given policy $\pi$, denoted as $v_{\pi}$, indicates how good any given state is for an agent following policy $\pi$. In other words, the value of state $s$ under policy $\pi$ is the expected return from starting from state $s$ at time $t$ and following $\pi$ thereafter. Mathematically, $v_{\pi}(s)$ is defined as:

\begin{equation}
v_{\pi}(s) = \mathbb{E}_{\pi}[G_{t}|S_{t}=s] = \mathbb{E}_{\pi} [\sum_{i=0}^{\infty} \gamma^{i} R_{t+i+1} | S_{t}=s]
\end{equation}

Similarly, the action-value function for a given policy $\pi$, denoted as $q_{\pi}$, indicates how good it is for the agent to take any given action from a given state while following policy $\pi$. In other words, the value of action $a$ in state $s$ under policy $\pi$ is the expected return from starting from state $s$ at time $t$, taking action $a$, and following policy $\pi$ thereafter. Mathematically, $q_{\pi}(s,a)$ is defined as:

\begin{equation}
q_{\pi}(s,a) = \mathbb{E}[G_{t}| S_{t} = s, A_{t} =a] = \mathbb{E}_{\pi} [\sum_{i=0}^{\infty} \gamma^{i} R_{t+i+1} |S_{t} = s, A_{t} = a ]
\end{equation}

It is also common to refer to the action-value function $q_{\pi}$ as the Q-function while the output from the Q-function for any given state-action pair (s,a) is referred to its Q-value. In other words, the Q-value associated with a state-action pair (s,a) represents the quality of taking action $a$ in state $s$.

As already defined, the goal of the agent is to derive a policy $\pi$ that maximizes the total amount of rewards received. That is, the policy that yields more return to the agent than all the other policies.

In terms of return, given two policies $\pi$ and $\pi^{'}$, policy $\pi$ is considered to be better than or the same as policy $\pi^{'}$ if the expected return of $\pi$ is greater than or equal to the expected return of policy $\pi^{'}$ for all states.
\begin{equation}
\pi \geq  \pi^{'} \; if\; and\; only\; if\; v_{\pi} (s) \geq  v_{\pi^{'}(s)}\; for\; all\; s \in S 
\end{equation}

A policy that is better than or at least the same as all other policies is called the optimal policy, and will be denoted $\pi_{*}$ from now on.

The optimal policy has an associated optimal state-value function, denoted as $v_{*}$ and defined as 
\begin{equation}
v_{*}(s) = \max_{\pi}^{} v_{\pi} (s)
\end{equation}
for all $s \in S$. In other words, $v_{*}$ gives the largest expected return achievable by any policy $\pi$ for each state.

Similarly, the optimal policy has an optimal action-value function, or optimal Q-function, which we denote as $q_{*}$ and define as 
\begin{equation}
q_{*}(s,a) = \max_{\pi}^{} q_{\pi}(s,a) 
\end{equation}
for all $s \in S$ and $a\in A(s)$. In other words, $q_{*}$ gives the largest expected return achievable by any policy $\pi$ for each possible state-action pair.

One fundamental property of $q_{*}$ is that it must satisfy the following equation.
\begin{equation}
q_{*}(s,a) = \mathbb{E}[R_{t+1}+ \gamma \max_{a^{'}}^{} q_{*}(s^{'},a^{'})]
\end{equation}
This is called the Bellman optimality equation. It states that, for any state-action pair (s,a) at time $t$, the expected return from starting in state $s$, selecting action $a$ and following the optimal policy $\pi_{*}$ thereafter is going to be the expected reward we get from taking action $a$ in state $s$, which is $R_{t+1}$, plus the maximum expected discounted return that can be achieved from any possible next state-action pair ($s^{'}$, $a^{'}$). Since the agent is following the optimal policy, the following state $s^{'}$ will be the state from which the best possible next action $a^{'}$ can be taken at time $t+1$

As a result, the optimal policy can be determined from the optimal state-value function $q_{*}$, because once $q_{*}$ has been solved, we can determine the optimal policy by finding the action $a$ that maximizes $q_{*}(s,a)$. Mathematically, this can be written as follows:
\begin{equation}
\pi_{*}(s) = \argmax_{a} q_{*}(s,a)
\end{equation}
In other words, the best action from any given state is the action that has the highest expected return based on the possible next states resulting from taking that action.

One algorithm to find the optimal Q-values for each state-action pair is the Q-learning algorithm~\citep{Watkins1992}. This algorithm iteratively updates the Q-values for each state-action pair using the Bellman equation until the Q-function converges to the optimal Q-function $q_{*}$. In the Q-learning algorithm, the Q-function is stored as a table with each cell corresponding to the Q-value of an action $a$ in a given state $s$. Then, the Q-values are updated iteratively according to the following update rule:
\begin{equation}
Q(s_{t}, a_{t}) =  (1 - \alpha)Q(s_{t}, a_{t}) + \alpha (r_{t}+ \gamma \max_{a}Q(s_{t+1}), a)
\end{equation}

Given infinite exploration time and a partly-random policy, it has been proved that the Q-learning algorithm can derive the optimal policy for any given finite MDP. However, although the Q-learning algorithm works well in very simple environments with a very limited number of actions and states such as the Frozen Lake~\footnote{\url{https://gym.openai.com/envs/FrozenLake-v0/}}, CartPole-v1~\footnote{\url{https://gym.openai.com/envs/CartPole-v1/}}, among others, this approach is totally impractical in complex environments as it requires having a finite state and action spaces and storing the full state-action table in memory (which is often unfeasible). As a result, the Double Q-learning algorithm~\citep{DBLP:journals/corr/MnihKSGAWR13,DBLP:journals/nature/MnihKSRVBGRFOPB15,10.5555/3016100.3016191} with experience replay~\citep{Lin1992} has been implemented to estimate the optimal Q-function. For a more detailed description of the algorithm, we refer the readers to Section~\ref{sec:agent}. 

Next, Section~\ref{sec:environment} describes the environment, and Section~\ref{sec:action_space} describes the action space. Afterwards, Section~\ref{sec:agent} describes the reinforcement learning algorithm used to approximate the optimal Q-function and the convolutional neural network architecture used to determine the best position to which insert the NOP instructions in order to bypass the malware classifier. Lastly, Section~\ref{sec:mlw_classifier} briefly resumes the machine learning classifier that we aim to bypass by modifying the assembly language source code of the executable with the insertion of NOP instructions.

\subsection{Environment}
\label{sec:environment}
The environment consists of an initial malware sample (one malware sample per episode) and the malware classifier (the attack target). Each time step or turn within an episode provides the following feedback to the agent:
\begin{itemize}
	\item A reward value $r_{t}\in \mathbb{R}$ given for mislabeling the malware family. The reward value $r_{t}$ at time step t is equal to the difference between the loss of the previous state and the loss of the current state of the classifier that we want to evade.
	\begin{equation}
	r_{t} = -1*(loss_{t-1}-loss_{t})
	\end{equation}
	where $loss_{t}$ refers to the multi-class logarithmic loss or cross entropy loss returned by the classification model for state $s_{t}$. Thus, if the reward at time step t, $r_{t}$, is positive, it indicates that the action performed at time step t, $a_{t}$, has increased the error loss of the target classifier. On the other hand, if the reward at time step t, $r_{t}$, is negative then, the action at time step t, $a_{t}$ has negatively contributed to the misclassification of the current executable.
	
	\item The state $s_{t}$ of the environment (malware sample) is represented as a sequence of assembly language instructions extracted from the assembly language source code of the Portable Executable file, whereas the assembly language source code can be obtained by disassembling the Portable Executable file using any disassembler of your choice, i.e. IDA Pro~\footnote{\url{https://www.hex-rays.com/ida-pro/}}, Radare2~\footnote{\url{https://rada.re/n/}}, Ghidra~\footnote{\url{https://ghidra-sre.org/}}. One particularity of using the sequence of instructions to represent an executable is that the resulting length of the sequence will differ from one executable to the other. In addition, for each instruction added to the executable, the size of the resulting sequence of instructions will be incremented by one. Instead of taking the assembly language source code as a whole, including its arguments, we decided to only take as input the mnemonics of the instruction. The mnemonic of a given instruction refers to the portion of the machine language instruction that specifies the operation to be performed, i.e. on encountering the instruction \emph{lea eax, [esp+8]},  we simply take as input the opcode \emph{lea}.
\end{itemize}
Based on the feedback provided by the reward, the agent learns which action to choose from a set of mutations (See Section~\ref{sec:action_space}) given the environment's state (sequence of assembly language instructions), while preserving the format and function of the PE file. 

\subsection{Action Space}
\label{sec:action_space}
\begin{figure}[ht]
	\centering
	\includegraphics[width=\columnwidth]{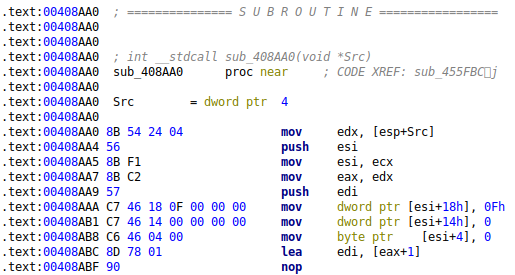}
	\caption{Snapshot of a piece of assembly language source code with a NOP instruction inserted at the location 0x00408ABF.}
	\label{fig:after_nop_insertion}
\end{figure}

The mutations represent the actions or moves available to the agent within the environment. Formally, the set of all possible actions is defined as $ A =\{insert_{0}, insert_{1},..., insert_{n}\}$, where $insert_{i}$ refers to inserting a~\emph{NOP} instruction at position $i$ in the assembly language instruction sequence. 
Therefore, at time $t$, the agent chooses action $a_{t}$. In this paper, we only considered as valid actions the insertion \emph{NOP} instructions as it is the simplest dead code instruction available~\citep{Balakrishnan05codeobfuscation}. The~\emph{NOP} or no-op instruction (short for no operation) is an assembly language instruction that does nothing. An example is provided in Figure~\ref{fig:after_nop_insertion}.
In addition, the number of actions available differs for each sample because it depends on the size and the content of their assembly language source code. Take into account that the locations to which the NOP instructions are inserted must not break the executable.

\subsection{Agent}
\label{sec:agent}
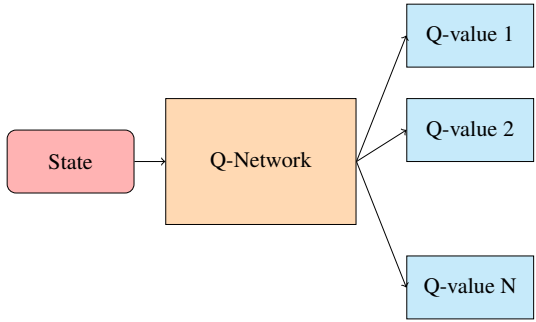
\begin{figure}[ht]
	\centering
	\usetikzlibrary{shapes.geometric, arrows.meta, backgrounds, fit, positioning}
	\usetikzlibrary{matrix,chains,scopes,positioning,arrows,fit}
	\tikzstyle{input} = [rectangle, rounded corners, minimum width=2cm, minimum height=1cm,text centered, draw=black, fill=red!30]
	\tikzstyle{network} = [rectangle, minimum width=3cm, minimum height=2cm, text centered, draw=black, fill=orange!30]
	\tikzstyle{qvalue} = [rectangle, minimum width=2cm, minimum height=1cm,text centered, draw=black, fill=cyan!20]
	\tikzstyle{arrow} = [thick,->,>=stealth]
	\resizebox{0.8\columnwidth}{!}{
		\begin{tikzpicture}
		
		\node (state) [input,xshift=-6cm] {State};
		\node (qnetwork) [network, right of=state,xshift=2cm] {Q-Network};
		\node (qvalue1) [qvalue, right of=qnetwork, xshift=2.3cm, yshift=2cm] {Q-value 1};
		\node (qvalue2) [qvalue, below of=qvalue1, yshift=-0.5cm] {Q-value 2};
		\node (qvalue3) [qvalue, below of=qvalue2, yshift=-1.5cm] {Q-value N};
		
		\draw[->] (state) -- (qnetwork);
		\draw[->] (qnetwork.east) -- (qvalue1.west);
		\draw[->] (qnetwork.east) -- (qvalue2.west);
		\draw[->] (qnetwork.east) -- (qvalue3.west);
		\end{tikzpicture}
	}
	\caption{Black-box representation of the Q-Network. The Q-Network takes as input the state, i.e. the assembly language source code of malware, and outputs the estimated Q-values of inserting a NOP instruction to all locations of the source code.}
	\label{fig:qnetwork}
\end{figure}

The agent takes the assembly language instructions of the sample and executes the action that will eventually yield the highest cumulative reward. This action is selected according to the Q-values provided by a deep Q-network (DQN). See Figure~\ref{fig:qnetwork}. For each given input state, the Q-network outputs the estimated Q-values for each action that can be taken from that state.

To train this network, we used the Double Q-learning algorithm~\citep{10.5555/3016100.3016191}, which decouples the action selection from the target Q-value generation, with experience replay~\citep{Lin1992}. The Double Q-learning algorithm uses two networks, the primary network, $Q$, and the target network, $Qtarget$, to select what is the best action to take for the next state and to calculate the target Q-value of taking that action at the next state, respectively. The weights of the primary network at time step $t$ are denoted as $\theta $ while the weights of the target network at time step $t$ are denoted as $\theta^{'}$. Notice that the primary network and the target network architectures are the same. The only difference between them is their weights. Mathematically, this can be formulated as follows:
\begin{equation}
\label{eq:targetQvalue}
Q_{*}(s_{t}, a_{t}) \approx r(s_{t}, a_{t}) + \gamma Qtarget(s_{t+1}, \argmax_{a} Q (s_{t+1}, a; \theta);\theta^{'})
\end{equation}

The update of the target network stays unchanged from the primary network, and it slowly copies the weights of the primary network $Q$ using the Polyak averaging:
\begin{equation}
\label{eq:polyak_avg}
	\theta^{'} \leftarrow \tau \theta + (1-\tau)\theta^{'} 
\end{equation}



To train Q-networks it is used a technique called experience replay during training~\citep{Lin1992}. The idea behind is to store the agent's experiences at each time step in a replay memory data set and use these experiences to update the weights of the Q-network through gradient descent. 

At a time step $t$, the agent's experience denoted as $e_{t}$ is defined as a tuple of four elements containing the state of the environment $s_{t}$, the action $a_{t}$ taken from state $s_{t}$, the reward $r_{t+1}$ the agent receives at time step $t+1$ as the result of performing action $a_{t}$ on the state $s_{t}$, and the next state of the environment $s_{t+1}$.
$$ e_{t} = (s_{t}, a_{t}, r_{t+1}, s_{t+1})$$
The advantages of experience replay are various. First, each experience might be potentially used in multiple weight updates allowing for greater data efficiency. Second, previous online learning approaches learned directly from consecutive samples with strong correlations between them. By sampling a subset of samples from the experience replay data set, these correlations are broken and the variance of the updates is reduced~\citep{DBLP:journals/corr/MnihKSGAWR13}.

In addition, to balance between exploration and exploitation during training, we use the epsilon-greedy action selection algorithm. This algorithm tackles the exploration-exploitation trade-off by taking an exploratory action with probability $\epsilon$ and a greedy action with probability $1-\epsilon$. Mathematically, the epsilon-greedy action selection algorithm   selects an action $a_{t}$ at time step $t$ as follows:
\begin{equation}
\label{eq:egreedy}
a_{t}= \left\{\begin{matrix}
\max Q_{t}(a) \;\;\; with \; probability\; 1-\epsilon \\
random\; action\;\;\; with \; probability \; \epsilon\\
\end{matrix}\right.
\end{equation}

In our case, the value of $\epsilon$ is set to $1.0$ at the beginning of the training, and it is continuously decreased at each time step until reaching $0.5$. This has been done to allow the agent to explore more often than not as the state and action spaces in our problem are very large.

For a complete description of the Double Q-learning algorithm, we refer the readers to Algorithm~\ref{alg:doubleQlearning}, the work of ~\cite{10.5555/3016100.3016191}, and the references therein.

\begin{algorithm}[ht]
	\caption{Double Q-learning algorithm~\citep{10.5555/3016100.3016191}}
	\label{alg:doubleQlearning}
	\KwIn{Initialize weights $\theta$ and $\theta^{'}$ of primary and target networks $Q$ and $Qtarget$, replay buffer D, $\tau <<1$}
	
	\For{each episode }
		{
			\For{each environment step}
			{
				Select action $a_{t}$ using Equation~\ref{eq:egreedy}
				
				Execute $a_{t}$ and observe next state $s_{t+1}$ and reward $r_{t}$
				Store transition $(s_{t}, a_{t}, r_{t+1}, s_{t+1})$ in replay buffer D
			}
			\For{each update step}
			{
				sample $e_{t} = (s_{t}, a_{t}, r_{t+1}, s_{t+1}) \sim  D$\\
				Compute target Q-value using Equation~\ref{eq:targetQvalue}
				
				Perform gradient descent step on 
				\begin{equation}
				(Q_{*}(s_{t}, a_{t})-Q (s_{t}, a_{t}))^{2}
				\end{equation}
				
				Update target network parameters using Eq.~\ref{eq:polyak_avg}
			}

		}
\end{algorithm}

\subsection{Q-Network Architecture}
An overview of the Q-network is described in Figure~\ref{fig:Qnetwork_detailed}. It consists of a convolutional layer to extract features from the sequence of assembly language instructions followed by a time-distributed layer that applies the same fully-connected layer to every time step.  The layers description is as follows:
\begin{figure*}[ht]
	\centering
	\includegraphics[width=\textwidth]{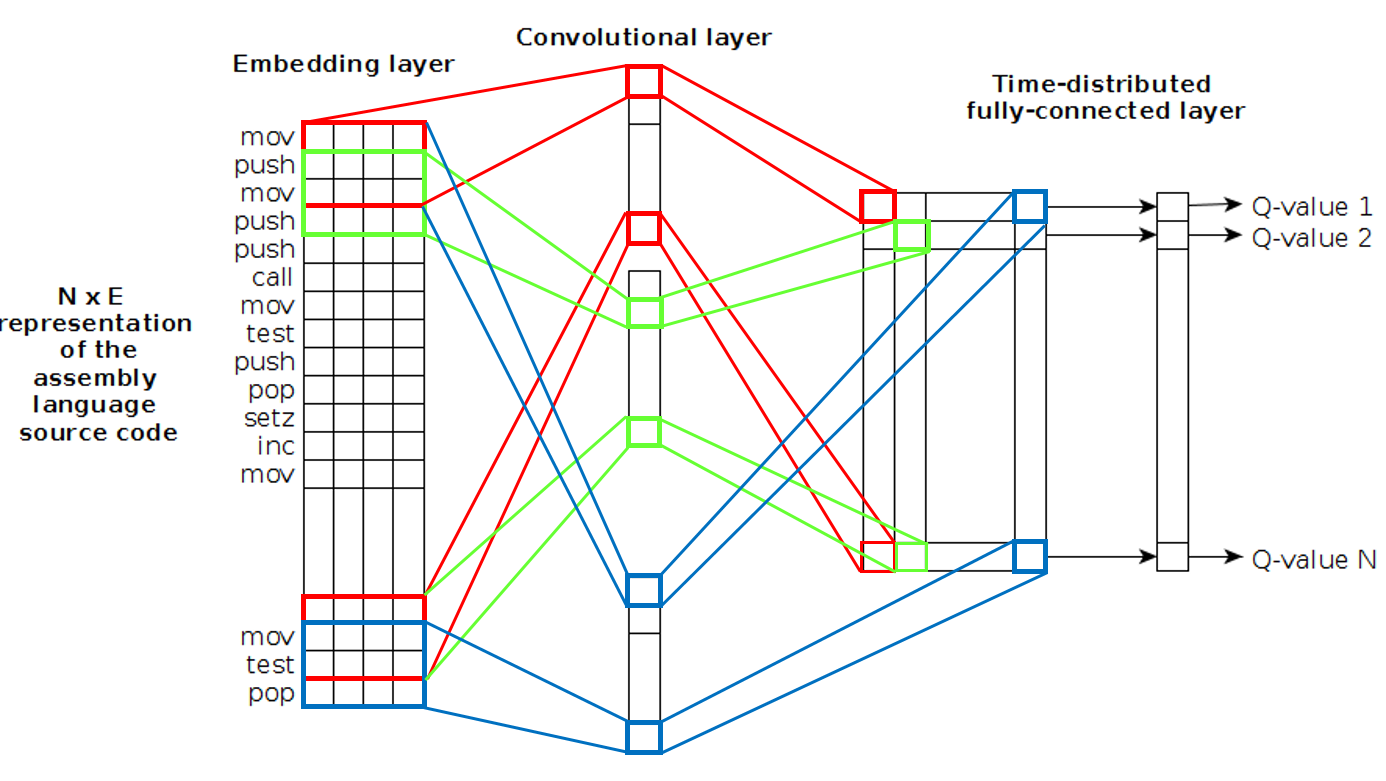}
	\caption{Graphical representation of the shallow convolutional Q-network architecture. The Q-network takes as input a Portable Executable (PE) file represented as a sequence of mnemonics (length N) and represents each mnemonic as a word embedding (size E). Afterwards, a convolutional layer with filters of size 3 is applied to extract features from the sequence of embedded mnemonics, followed by a time-distributed layer to retrieve the Q-value of inserting a NOP instruction at every location in the mnemonics sequence.}
	\label{fig:Qnetwork_detailed}
\end{figure*}
\begin{itemize}
	\item Input layer. The network takes as input an executable represented as a sequence of mnemonics. A mnemonic is the name of the operation a machine can execute. For instance, the assembly language instruction \emph{mov ebp, esp} is reduced to the mnemonic \emph{mov}. The main argument behind this simple representation is that it will generalize better as it would not be affected by small permutations in the arguments and thus, the obfuscation technique known as register reassignment would not alter the output of the classifier. 
	\item Embedding layer. Each mnemonic is represented as a low-dimensional vector of real values (word embedding) of size $E$, where each value captures a dimension of the mnemonics' meaning. $E$ was set to 4 as in~\citep{DBLP:conf/ccia/GibertBMPSV17}. The rationale behind using distributed representations is to better capture the semantic relationships between comparable mnemonics.
	\item Convolutional layer. The convolutional layer extracts N-gram like features from the sequence of assembly language instructions. The size of each filter is $h\; x\; E$ where $h=3$. The number of different filters in the convolutional layer is 10.
	\item Time-Distributed layer. This layer applies a fully-connected layer to each of the $N$ mnemonics ($N$ is equal to the size of the assembly language instructions sequence), independently. The input of the fully-connected layer is equal to the number of filters in the convolutional layer while the number of output neurons $o$ is equal to 1. 
	At the end, the softmax function is applied to output the estimated Q-values of inserting a NOP instruction in any location of the assembly language instructions given as input to the network.  
\end{itemize}
For a complete description of the trainable parameters of the Q-network we refer the readers to Table~\ref{tab:qnetwork_parameter_details}.

\subsection{CNN classifier}
\label{sec:mlw_classifier}
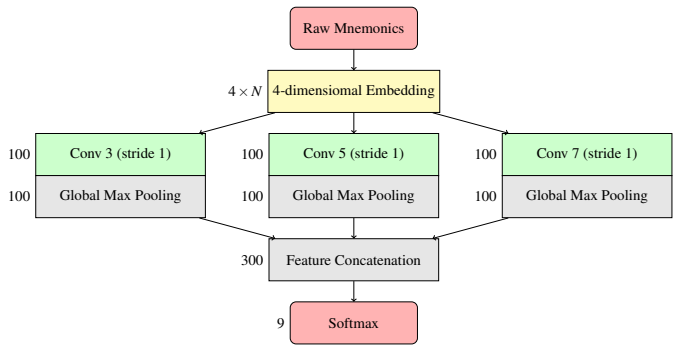
\begin{figure}[ht]
	\centering
	\tikzstyle{state}=[circle,
	thick,
	minimum size=1.2cm,
	draw=black,
	fill=white]
	\tikzstyle{input_output} = [rectangle, rounded corners, minimum width=3cm, minimum height=1cm,text centered, draw=black, fill=red!30]
	\tikzstyle{relu_act} = [rectangle, minimum width=1cm, minimum height=1cm,text centered, draw=black, fill=green!20]
	\tikzstyle{softmax} = [rectangle, minimum width=3cm, minimum height=1cm,text centered, draw=black, fill=red!30]
	\tikzstyle{selu} = [rectangle, minimum width=4cm, minimum height=1cm,text centered, draw=black, fill=blue]
	\tikzstyle{relu} = [rectangle, minimum width=4cm, minimum height=1cm,text centered, draw=black, fill=green!20]
	\tikzstyle{nonactivation} = [rectangle, minimum width=3cm, minimum height=1cm,text centered, draw=black, fill=white]
	
	\tikzstyle{embedding} = [rectangle, minimum width=4cm, minimum height=1cm,text centered, draw=black, fill=yellow!30]
	\tikzstyle{pooling} = [rectangle, minimum width=4cm, minimum height=1cm,text centered, draw=black, fill=gray!20]
	\resizebox{\columnwidth}{!}{%
		
		\begin{tikzpicture}[node distance=1.5cm]
		\node (raw_mnemonics) [input_output] {Raw Mnemonics};
		\node (emb) [embedding, below of=raw_mnemonics, label={left:$4 \times N$}] {4-dimensiomal Embedding};
		\node (conv_4) [relu, below of=emb, label={left:$100$}] {Conv 5 (stride 1)};
		\node (conv_3) [relu, left of=conv_4, label={left:$100$}, xshift=-4cm] {Conv 3 (stride 1)};
		\node (conv_5) [relu, right of=conv_4, label={left:$100$}, xshift=4cm] {Conv 7 (stride 1)};
		
		
		\node (pooling4) [pooling, below of=conv_4, label={left:$100$}, yshift=0.5cm] {Global Max Pooling};
		\node (pooling3) [pooling, below of=conv_3, label={left:$100$}, yshift=0.5cm] {Global Max Pooling};
		\node (pooling5) [pooling, below of=conv_5, label={left:$100$}, yshift=0.5cm] {Global Max Pooling};
		\node (concat) [pooling, below of=pooling4, label={left:$300$}] {Feature Concatenation};
		\node (softmax) [input_output, below of=concat, label={left:$9$}] {Softmax};
		
		\draw[->] (raw_mnemonics) -- (emb);
		\draw[->] (emb) -- (conv_4);
		\draw[->] (emb) -- (conv_3);
		\draw[->] (emb) -- (conv_5);
		\draw[->] (pooling4) -- (concat);
		\draw[->] (pooling3) -- (concat);
		\draw[->] (pooling5) -- (concat);
		\draw[->] (concat) -- (softmax);
		
		\end{tikzpicture}
	}
	\caption{State-of-the-art convolutional neural network architecture for malware classification~\citep{DBLP:conf/ccia/GibertBMPSV17}.}
	\label{fig:cnn_classifier}
\end{figure}

The classifier model that our agent is seeking to evade is a shallow convolutional neural network specifically designed to learn N-gram based features from a sequence of text~\citep{DBLP:conf/ccia/GibertBMPSV17,GIBERT2021102159}. This classifier has been selected among the deep learning classifiers in the literature because of its state-of-the-art performance~\citep{GIBERT2021102159}. See Figure~\ref{fig:cnn_classifier} and Table~\ref{tab:shallow_cnn_malware_classifier_parameter_details} for a detailed description of the network architecture.  

The network takes as input an executable represented as a sequence of mnemonics and convolves various filters of different sizes to extract n-gram like features. In particular, the filters have sizes 3, 5 and 7. Afterwards, a global max-pooling layer is applied to extract the maximum activation of each of the feature map activations from the previous layer. At the end, the softmax function is applied to the linear combination of the learned features to output the probability distribution over malware families.

\section{Evaluation}
\label{sec:evaluation}
\subsection{The Microsoft Malware Classification Dataset}
\begin{figure}[ht]
	\centering
	\includegraphics[width=\columnwidth]{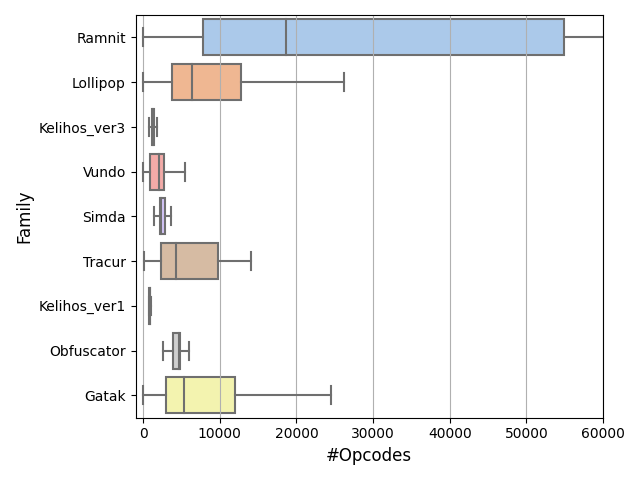}
	\caption{Average number of instructions per family. It can be observed that the size of the samples belonging to different families is not similar. For instance, the samples belonging to the Ramnit and Lollipop families contain $46601.44$ and $19157.06$ instructions per sample, respectively, while the samples belonging to the Kelihos\_ver1 family have only an average of $1326.65$ instructions per sample.}
	\label{fig:opcodes_per_class}
\end{figure}

To validate our method, the data provided by Microsoft for the  Big Data Innovators Gathering~\citep{DBLP:journals/corr/abs-1802-10135} has been used instead of creating our own in-house (non-reproducible) dataset of benign and malicious samples for reproducibility purposes. Notice that, unlike other domains, there are legal restrictions in place that forbid sharing benign binaries such as copyright laws. In addition, determining whether a file is malicious and its corresponding family is very time-consuming even for experienced security analysts.
Furthermore, the Microsoft dataset has become the standard benchmark to evaluate the performance of machine learning algorithms for the task of Windows malware classification because it provides high quality samples from various malware families.
The dataset is publicly accessible and hosted on Kaggle~\footnote{https://www.kaggle.com/c/malware-classification/overview}, and it contains almost half of a terabyte of malware belonging to 9  malware families. See Table~\ref{tab:class_distribution}.
\begin{table}[ht]
	\centering
	\caption{Class distribution in the Microsoft dataset~\citep{DBLP:journals/corr/abs-1802-10135}.}
	\label{tab:class_distribution}
	\begin{tabular}{lll}
		\hline
		Family Name & \# Train Samples & Type \\ \hline
		Ramnit & 1541 & Worm \\
		Lollipop & 2478 & Adware \\
		Kelihos\_ver3 & 2942 & Backdoor \\
		Vundo & 475 & Trojan \\
		Simda & 42 & Backdoor \\
		Tracur & 751 & TrojanDownloader \\
		Kelihos\_ver1 & 398 & Backdoor \\
		Obfuscator.ACY & 1228 & Obfuscated malware \\
		Gatak & 1013 & Backdoor \\ \hline
	\end{tabular}
\end{table}
For each file, it is provided the hexadecimal representation of the file's binary content and the assembly language source code generated with the IDA disassembler tool~\footnote{https://www.hex-rays.com/products/ida/}. The assembly language source code of a computer program is the low-level representation of the program's statements and machine code instructions. As observed in Table~\ref{tab:class_distribution} and Figure~\ref{fig:opcodes_per_class}
, the dataset is imbalanced and heterogeneous, with the average number of assembly language instructions of samples belonging to different families very distinct.
This particularity will greatly affect the performance of the framework as shown in Section~\ref{sec:results}.

Notice that the methodology described through this paper can be applied for the task of malware detection with minor modifications, i.e. reducing the number of output neurons from 9 (number of families in the Microsoft dataset) to 1, indicating the maliciousness of the executable.  
\subsection{Experimental Setup}
\label{sec:experimental_setup}
The experiments were run on a computer with the following specifications: Intel i7-7700K, 32 GB RAM, 2 x Nvidia GTX 1080Ti. The Microsoft dataset comprises two sets, the training and the test set. But unfortunately, the labels of the test set are not provided. To evaluate the models on the test set, a file with the predicted class probabilities for each sample on the set must be submitted on Kaggle. In consequence, to evaluate our framework, we divided the training set into three sets: training, validation and test set, each containing  70\%, 15\% and 15\% of the samples of the original training set, respectively. 

\subsection{Parameters Setting}
\begin{figure}[ht]
	\centering
	\includegraphics[width=0.95\columnwidth]{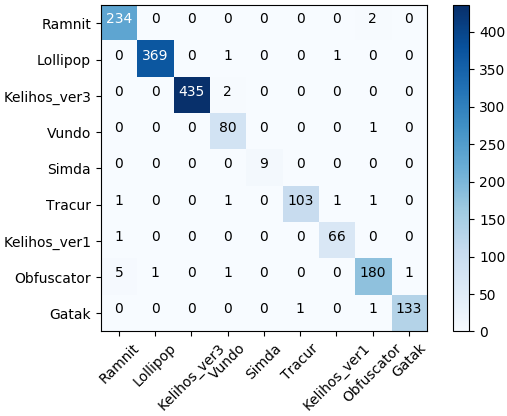}
	\caption{Confusion or error matrix of the CNN classifier on the test set. Each row represents a predicted class and each column represents the instances in an actual class. By definition, a confusion matrix $C$ is such that $C_{i,j}$ is equal to the number of observations known to be in family $i$ and predicted to be in family $j$. The diagonal represents the case where the prediction of the CNN model is family $i$ and the actual class is $i$ too. Any off-diagonal entry indicates some mistake.}
	\label{fig:cnn_confusion_matrix}
\end{figure}

In the experiments, the malware classifier to be evaded is trained on the training set for 15 epochs and evaluated on the validation set. The test set is then used to provide an unbiased evaluation of the fitness of the final model. Figure~\ref{fig:cnn_confusion_matrix} shows the confusion matrix, also known as the error matrix, which summarizes the results of testing the model on the test set. The accuracy achieved is 98.65\%. Regarding the hyperparameters of the classifier, the convolutional layer has 300 filters of size $h\; x\; M$, where $h \in \{3,5,7\}$ and $M$ is the embedding size ($M\; =\;4$).

For the reinforcement learning framework parameters setup, the maximum number of modifications per episode is $50$ to limit the number of modifications to the minimum. As observed in Figure~\ref{fig:average_steps_per_class}, the average number of dead code insertions needed to degrade the performance of the classifier using a random agent on average is less than 50 for the Gatak, Kelihos\_ver1, Tracur, Simda, Vundo and Kelihos\_ver3 families. Thus, we considered that 50 insertions were more than enough to test whether or not the AI agent outperforms the random agent. For each family, a different model was trained for a total number of $1000$ episodes. In consequence, each model learns which N-grams need to break or mimic in order to mislabel the samples of a given malware family. The discount factor $\gamma$ is set to $0.99997$. The major parameter settings in the training algorithm are shown in Table~\ref{tab:qnetwork_parameters}.
\begin{table*}[ht]
	\centering
	\caption{List of major parameters and their values in the training algorithm.}
	\label{tab:qnetwork_parameters}
	\resizebox{\textwidth}{!}{%
		\begin{tabular}{l|l|l}
			\hline
			Parameter settings            & Value   & Description                                                                                        \\ \hline
			MAXTURN                       & 50      & The maximum number of modifications allowed to be performed by the agent before declaring failure. \\
			EPISODES                      & 1000   & The maximum number of episodes played for each family.                                             \\
			Memory buffer max size        & 2000    & The capacity of the replay buffer memory.                                                          \\
			Discount factor               & 0.99997 & The discount factor $\gamma$ used in the Q-learning update.                                          \\
			Initial exploration           & 1       & Initial value of the $\epsilon$ in $\epsilon$-greedy exploration                                       \\
			Final exploration             & 0.5     & Final value of the $\epsilon$ in $\epsilon$-greedy exploration                                         \\
			Number of filters (Q-Network) & 10      & Number of filters in the convolutional layer of the Q-Network.                                     \\ \hline
		\end{tabular}%
	}
\end{table*}
\subsection{Results}
\label{sec:results}
\begin{figure}[ht]
	\centering
	\includegraphics[width=\columnwidth]{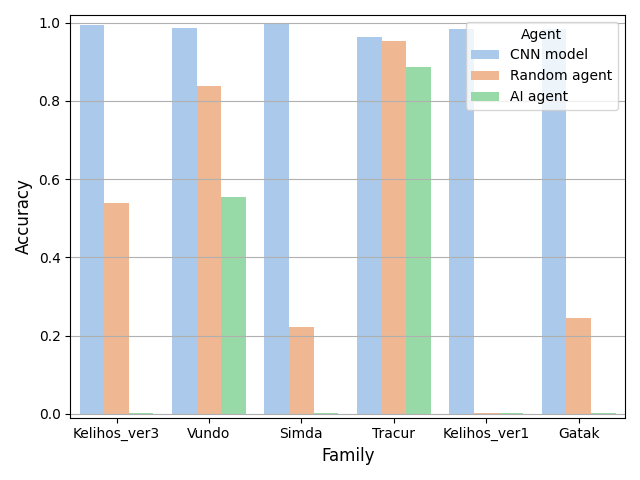}
	\caption{Performance comparison of the CNN model on the test set. It can be seen that the DQN agent outperforms the random agent in almost all families (The lower is the accuracy of the model on the resulting obfuscated test set, the better). Notice that those families in which the agents haven't succeeded in misclassifying any of the samples are not displayed, i.e. Ramnit, Lollipop and Obfuscator.ACY families.}
	\label{fig:accuracy_per_class}
\end{figure}

\begin{figure}[ht]
	\centering
	\includegraphics[width=\columnwidth]{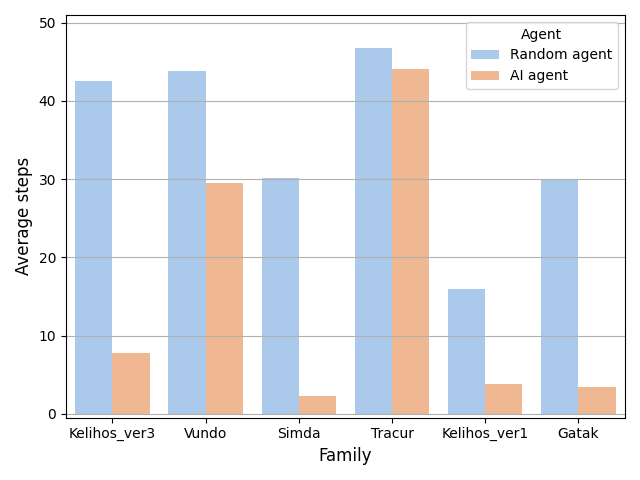}
	\caption{Comparison of the average number of NOP instructions required by the DQN and random agents to misclassify the state-of-the-art CNN classifier~\citep{DBLP:conf/ccia/GibertBMPSV17,GIBERT2021102159}. Notice that those samples in which the agents haven't succeeded in misclassifying any of the samples are not displayed, i.e. Ramnit, Lollipop and Obfuscator.ACY. In those cases, the average number of NOP insertions is 50, which is equal to \emph{MAXTURN}, the maximum number of modifications allowed to be performed by the agent before declaring failure.}
	\label{fig:average_steps_per_class}
\end{figure}

\begin{figure}[ht]
	\centering
	\includegraphics[width=\columnwidth]{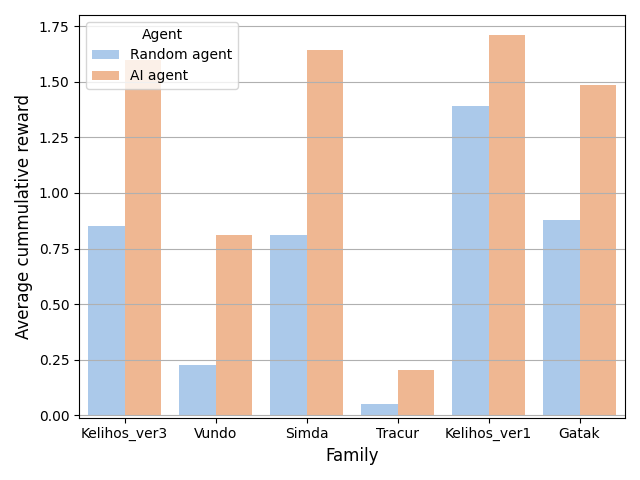}
	\caption{Comparison of the average cumulative reward achieved by the DQN and random agents. Notice that those families in which the agents haven't succeeded in misclassifying any of the samples are not displayed as their average cumulative reward is almost 0, i.e. Ramnit, Lollipop and Obfuscator.ACY families.}
	\label{fig:reward_per_class}
\end{figure}

The reinforcement learning models are trained on the samples from the validation set. To evaluate the performance of the learned models in mislabeling the malicious samples of any given family, we recorded the average total reward value, the average number of NOP insertions and the average accuracy per family on the test set. The average total reward plot in Figure~\ref{fig:reward_per_class} shows a 87.62\%, 260.95\%, 103.09\%, 288.05\%, 22.88\%, 68.75\% increase with respect to the cumulative rewards achieved by a random agent on the Kelihos\_ver3, Vundo, Simda, Tracur, Kelihos\_ver1 and Gatak malware families, respectively. In addition, as it can be observed in Figures~\ref{fig:accuracy_per_class} and~\ref{fig:average_steps_per_class}, the AI agent was able to misclassify all the samples on the test set belonging to the Kelihos\_ver3, Simda, Kelihos\_ver1 and Gatak families while also reducing by 81.81\%, 92.62\%, 76.47\%, and 88.40\% the number of NOP instructions inserted, respectively.
As evidenced by the experiments (Table~\ref{tab:results}), our framework is able to generate adversarial examples that bypass a state-of-the-art CNN malware classifier by only inserting NOP instructions. However, the agent has problems mislabeling the samples belonging to the Ramnit, Lollipop, Tracur and Obfuscator.ACY families because of the following two reasons: First, the samples of malware belonging to the Kelihos\_ver3, Simda, Kelihos\_ver1 and Gatak families in the training set do not contain any NOP instruction. As a result, their respective Q-networks ended up learning the optimal locations to where to insert a NOP instruction in such a way that it makes the malware classifier detect a pattern characteristic of samples in a different malware family. On the contrary, the malware samples belonging to the Ramnit and Lollipop families hardly break the patterns the classifier has learned to detect as their assembly language instructions are at least twice as bigger than those of the other families, and some patterns learned for those families consist of at least one NOP instruction. You can observe in Table~\ref{tab:results} that inserting 50 NOP instructions is not enough to misclassify the executables belonging to these families. This could be solved by increasing the maximum number of mutations to be performed or by allowing the agent to insert other dead instructions such as (1) MOV Reg, Reg, (2) PUSH Reg; POP Reg, (3) ADD Reg, 0, to see if with the insertion of those instructions we could make the agent mimic patterns learned for other families. 

Nevertheless, in the present paper, we have demonstrated that the use of simple obfuscation techniques is more than enough to decrease the overall accuracy of deep learning models, and we have provided a general framework that uses reinforcement learning to boost the efficiency of the dead code insertion technique. 

\begin{table*}[ht]
	\centering
	\caption{Comparison of the accuracy of the CNN model on the samples of the test set and the obfuscated versions generated by the random agent and the DQN agent, as well as the average cumulative reward and the average number of NOP insertions of both agents.}
	\label{tab:results}
	\begin{tabular}{l|lll|ll|ll}
		\hline
		& \multicolumn{3}{l|}{CNN accuracy}   & \multicolumn{2}{l|}{Average number of NOP insertions} & \multicolumn{2}{l}{Average cummulative reward} \\ \hline
		Malware family & No agent & Random agent & DQN agent & Random agent                & DQN agent               & Random agent             & DQN agent            \\ \hline
		Ramnit         & 99.15    & 99.15        & 99.15     & 48.58                       & 48.58                   & 0.01                     & 0.01                 \\
		Lollipop       & 99.46    & 99.46        & 99.19     & 48.74                       & 48.612                  & 0.01                     & 0.01                 \\
		Kelihos\_ver3  & 99.54    & 53.78        & 0.0       & 42.60                       & 7.75                    & 0.85                     & 1.60                 \\
		Vundo          & 98.77    & 83.95        & 55.56     & 43.79                       & 29.47                   & 0.22                     & 0.81                 \\
		Simda          & 100.0    & 22.22        & 0.0       & 30.11                       & 2.22                    & 0.81                     & 1.64                 \\
		Tracur         & 96.26    & 95.33        & 88.79     & 46.84                       & 44.14                   & 0.05                     & 0.21                 \\
		Kelihos\_ver1  & 98.51    & 0.0          & 0.0       & 15.99                       & 3.76                    & 1.39                     & 1.71                 \\
		Obfuscator.ACY & 95.74    & 95.21        & 95.74     & 46.85                       & 46.95                   & -0.01                    & 0.01                 \\
		Gatak          & 98.52    & 24.44        & 0.0       & 29.83                       & 3.46                    & 0.88                     & 1.48                 \\ \hline
	\end{tabular}
\end{table*}

\section{Conclusions}
\label{sec:conclusions}
This paper proposes a general framework using reinforcement learning to make a state-of-the-art malware classifier to incorrectly label the samples of a given family. The core component is an intelligent agent, which constantly interacts with malware samples to learn to choose the optimal locations to where to insert NOP instructions. This has been achieved by training a shallow convolutional Q-network using the double Q-learning algorithm. Experiments show that the proposed framework dropped the classification accuracy of the classifier from 98.65\% to 56.53\% on the test set while having an evasion rate of 100\% for the samples belonging to the Kelihos\_ver3, Simda, and Kelihos\_ver1 families. In addition, the Q-network enhanced the evasion rate of the dead code insertion technique by  11.29\% while also reducing the average number of NOP insertions needed to mislabel the malware sample by  33\%.

\subsection{Future Work}
One future line of research is the use of additional dead code instructions (such as the \emph{MOV Reg, Reg} instruction, the \emph{ADD Reg, 0} instruction, the \emph{SUB Reg, 0} instruction, the \emph{SHL Reg, 0} instruction, among others), and the investigation of other common obfuscation techniques such as the instruction substitution technique, the subroutine reordering technique, and the code reordering through jumps technique. Moreover, instead of developing adversarial attacks, one line of research could be the study of potential methods for hardening anti-malware engines against adversarial crafting.

\section*{Acknowledgements}
This project has received funding from the European Union’s Horizon 2020 research and innovation programme under the Marie Skłodowska-Curie grant agreement No. 847402. This research has been partially funded by the Spanish MICINN Projects TIN2015-71799-C2-2-P, ENE2015-64117-C5-1-R, PID2019-111544GB-C22, and supported by the University of Lleida.

\bibliographystyle{elsarticle-harv.bst}
\bibliography{refs}

\appendix
\section{Configuration Details of the Neural Networks}

\begin{table}[ht]
	\centering
	\resizebox{\columnwidth}{!}{%
		\begin{tabular}{lll}
			\hline
			Layer (type)                           & Output shape   & Parameters \# \\ \hline
			input\_1 (Input layer)                 & (None, N, 1)   & 0             \\
			embedding\_1 (Embedding layer)         & (None, N, E)   & 4*557 (E*V)      \\
			conv1d\_3(Conv\_1D)                    & (None, N, 10)  & 10*3*4         \\
			global\_maxpool1d\_3 (GlobalMaxPooling1D) & (None, 10)     & 0             \\
			time\_distributed (dense\_1)    & (None, N)      & 10*1 \\
			softmax\_1 (Softmax)                   & (None, N)      & 0             \\ \hline
			\multicolumn{2}{l}{Total trainable parameters}          & 2358             \\ \hline
		\end{tabular}%
	}
	\caption{Configuration details of the Q-network architecture used to select the locations to which insert NOP instructions in a given malware sample in order to bypass detection. N is the size of the input mnemonics sequence. E is the embedding size. V is the vocabulary size, i.e. the number of different mnemonics found in the Microsoft dataset.}
	\label{tab:qnetwork_parameter_details}
\end{table}

\begin{table}[ht]
	\centering
	\resizebox{\columnwidth}{!}{%
		\begin{tabular}{lll}
			\hline
			Layer (type)                           & Output shape   & Parameters \# \\ \hline
			input\_1 (Input layer)                & (None, N, 1)   & 0             \\
			embedding\_1 (Embedding layer)            & (None, N, E)   & 4*557 (E*V)   \\
			conv1d\_3(Conv\_1D)                    & (None, N, 100) & 100*3*4         \\
			global\_maxpool1d\_3 (GlobalMaxPooling1D) & (None, 100)    & 0             \\
			conv1d\_5(Conv\_1D)                    & (None, N, 100) & 100*5*4         \\
			global\_maxpool1d\_5 (GlobalMaxPooling1D) & (None, 100)    & 0             \\
			conv1d\_7(Conv\_1D)                    & (None, N, 100) & 100*7*4         \\
			global\_maxpool1d\_7 (GlobalMaxPooling1D) & (None, 100)    & 0             \\
			features concatenation                 & (None, 300)    & 0             \\
			dense\_1 (Dense)                       & (None, 9)      & 300*9         \\
			softmax\_1 (Softamx)                  & (None, 9)      & 0             \\ \hline
			\multicolumn{2}{l}{Total trainable parameters}          & 10928             \\ \hline
		\end{tabular}%
	}
	\caption{Configuration details of the shallow CNN architecture for malware classification.}
	\label{tab:shallow_cnn_malware_classifier_parameter_details}
\end{table}

\end{document}